# The role of cost in the integration of security features in integrated circuits for smart cards

Nikolaos Athanasios Anagnostopoulos
**Student number**: 1318055

**Research Topics**

University of Twente

EIT ICT Labs Master School

2013-14

UNIVERSITEIT TWENTE.

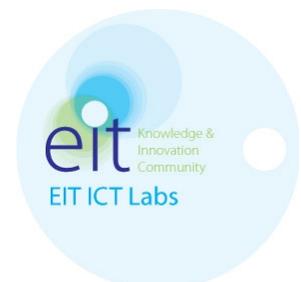

# Abstract

This essay investigates the role of cost in the development and production of secure integrated circuits. Initially, I make a small introduction on hardware attacks on smart cards and some of the reasons behind them. Subsequently, I introduce the production phases of chips that are integrated to smart cards and try to identify the costs affecting each one of them. I proceed to identify how adding security features on such integrated circuits may affect the costs of their development and production. I then make a more thorough investigation on the costs of developing a hardware attack for such chips and try to estimate the potential damages and losses of such an attack. I also go on to examine potential ways of reducing the cost of production for secure chips, while identifying the difficulties in adopting them.

This essay ends with the conclusion that adding security features to chips meant to be used for secure applications is well worth it, because the costs of developing attacks are of comparable amounts to the costs of developing and producing a chip and the potential damages and losses caused by such attacks can be way higher than these costs. Therefore, although the production and development of integrated circuits come at a certain cost and security introduces further additional costs, security is inherently unavoidable in such chips. Finally, I additionally identify that security is an evolving concept and doesn't aim to make a chip totally impenetrable, as this may be impossible, but to lower the potential risks, including that of being compromised, to acceptable levels. Thus, a balance needs be found between the level of security and the levels of cost and risk.

# PART I

## General introduction

Smart cards are becoming really popular for secure transactions and identification purposes. This situation is naturally inclined to introduce higher incentives for both culprits and researchers to discover and commence attacks against the integrated circuits of such cards. In turn, these intense efforts to circumvent the security of such electronic chips and gain access to their confidential data result in a raising demand for the integration of better and more efficient security features in such chips.

As I work for a semiconductor company manufacturing secure chips for smart cards, it is really important for it to be able to assess how adding security features to a chip may affect its development and production costs. Additionally, the company needs to examine whether potential losses and damages caused by attacks on such chips can justify this increased cost caused by the integration of security features in its chips.

For example, NXP's MIFARE brand of smart cards has recently come under increased attack in an effort to identify potential ways to overcome its security features and gain access on confidential information stored on the cards.[1][2][3][4][5][6][7][8][9] Researchers have quite often succeeded in compromising the security of these smart cards.[1][2][3][6][7][9] The

MIFARE company claims to possess "a confirmed market share of 77% in public transport",[10] which is equivalent to at least 1 billion of its cards being used for secure online transactions related to transportation.[6] This means that if the techniques demonstrated by these researchers were to be employed by culprits, and assuming that around 1% of the cards used were to be exploited, this could potentially result in losses amounting to millions of euros in a very short period of time.

Apart from direct economic losses and costs, a hardware attack on the integrated circuits of smart cards can also cause serious harm to the relevant company's prestige and brand name. Potential future clients will be quite unwilling to place their trust on a company that has failed to secure its current clients' private data and their online – or offline – transactions. Therefore, it seems extremely crucial for a company manufacturing chips for smart cards to place significant effort and resources on assuring their security, by strengthening any security measures that have already been implemented on them and developing further more adequate and efficient ways to reinforce their security.

It is also really important for manufacturers of chips for smart cards to continuously monitor the most recent developments in attacking such integrated circuits, because only in this way an adequate effort can be made towards providing sufficient countermeasures against the latest reported attacks. In general, because the development of a new chip usually takes at least a couple of years, it is inevitable that new ways of compromising the chip's security measures will be developed before the chip reaches the market.[11] As this new information cannot be integrated in a chip's architecture, design and features, while this is already being developed, without considerable new cost, the original design of such an integrated circuit must already implement a substantial degree of novelty regarding its security features, which may be able to counter future attacks.[11]

It also evident that any such new design for a new chip to be used in smart cards needs to be extensively tested, as it regards applications which require a significantly high degree of security. A new integrated circuit solution for such a purpose not only has to work efficiently, but also must sufficiently protect its confidentiality and integrity against not only already known, but also potential future, attacks. Therefore, it requires an exhaustive amount of designing and testing both as a prototype, but also when manufactured in mass quantities.

Thus, although the integration of security measures in such chips is intentionally sought-after and inherently unavoidable,[12] it comes at a large cost. Inherent costs exist due to the research and development phase of such features, which include manpower and infrastructure costs, as well as costs of design and implementation for the prototypes of the new security features and the whole of the chip, as the chip may need to be significantly redesigned. Furthermore, significant costs may also occur during the mass production of the chip, as the new features may increase the chip's area and thus raise the cost of the materials required for its production. Moreover, the addition of security features may also introduce performance costs, related to the chip's computation time, delay costs, as well as power costs, due to the increased complexity of the new integrated circuit and the need to also provide enough power for the additional features. Additionally, the integration of such novel security features can also imply a rise in packaging and marketing costs and surely involves a significant rise in costs for testing, as the new features require unit testing on their own, integration testing towards the rest of the chip and a full system and acceptance testing for the new chip as a whole.

Finally, it should be noted that even though a hardware attack could potentially result in significant losses and damages, it usually also requires a higher level of knowledge and equipment than a software attack. Furthermore, such attacks can far less often be automated in comparison to software attacks, while they quite often require the use of advanced equipment in a very specific way and expert knowledge, not only in the general field, but also about the very specific hardware to come under attack. Moreover, as hardware attacks are more physical in their nature than software attacks, they often require a large number of integrated circuits of the same kind in order to be successful, which may generally be hard to obtain. In conclusion, this means that performing a hardware attack tends to have a high cost, which may or may not be justified by its potential benefits and rewards for the attacker.

# The business case to be examined: Current state of costs related to development and production of secure chips for smart cards

As I work for a company manufacturing secure chips for smart cards, it is important to examine how adding security features to a chip may affect its costs. For obvious reasons, such integrated circuits embedded on the surface of these cards require a high level of security to protect the confidentiality, integrity and availability of their data. It has been noted that smart cards which are used primarily for secure applications need to have a variety of security features implemented on their chips and, thus, they employ specifically developed integrated circuits.[12]

For this purpose, we need to explore the way in which computer chips are generally produced and identify the changes that implementing extra security features may cause to this process. We can then examine the effects that the integration of secure features may have towards cost in each stage of a chip's development from research to the market. Consequentially, we may be able to suggest possible ways in which this cost may be reduced and/or the process be optimised, without significantly affecting the quality of the end product.

It is therefore important to determine the level of cost each step introduces in the process of development of a common chip towards its total cost and/or identify the specific costs associated with that stage. Then, we can examine how implementing security features on such an integrated circuit may affect these costs or the overall level of that stage's cost. And, finally, we should try to detect the different methods and means which can minimise each step's cost and/or the overall cost of the final product, an integrated circuit designed for smart cards.

To this end, it may be useful to state the following research questions, which this essay will try to address:

- What is the amount of additional cost that integrating security features on a chip may introduce?
- Are security costs related to the costs of performing an attack and the potential damages and losses caused by them?
- How can we decrease this amount of additional cost introduced by (additional) security?
- Is it justified to introduce security on a chip used for secure transactions, and to what extent is this feasible?

# PART II

## The phases of production of a chip

In order to be able to calculate the costs of development and production for any chip, it is essential to properly classify the different stages of development and production. To this end, I have identified the following stages as essential for the development and production of a chip:

- Initial research conducted based on improving a previous product or on a new idea or design. This stage requires an initial investment of various resources.
- More in-depth research accompanied with initial or intermediate-level development of some designs of the chip. These designs should result in one or more prototypes per design. The first prototype should be a proof-of-concept one and the final a quite functional one, based on well-researched designs.
- Testing performed on the prototypes designed in the previous step and assessment of their functionality, quality and of the advantages and disadvantages of each design.
- Full development of one or more selected designs, resulting in entirely working prototypes which will serve as initial guides for the final selection of a single design to be put into mass production.
- More thorough testing of the initially selected designs and their related prototypes and subsequently really exhaustive testing and assessment of the selected final design and prototype to be used for mass production.
- Mass production of a product, which in some cases may be preceded by a phase of "experimental" small scale production serving as a last safety measure against potential failures.
- Initial testing of the functionality and quality of every massively produced series of the product, followed by more thorough testing as the product enters the market and is slowly accepted in it.
- Marketing of the product and subsequent small scale improvements of its design

and/or placement in the market, together with optimisation of the production process and gradual improvement of the product's quality.

It may be evident that the present essay is in itself a very partial implementation of the last stage of the production of a chip as described above. This fact, in turn, could present to us exactly how that stage of development may involve cost in it, as this very final stage may involve the production of a number of focused reports and essays along with targeted investigation and analysis of the previous stages of development, and all these activities of course involve costs in terms of time, manpower, knowledge and money.

## Agents of cost involved in the different stages of a chip's development and production

Several different factors may add cost in each of the stages of the development of a chip which I have previously identified. It is however essential to try to identify these elements of cost in each different step of development, in order to provide efficient ways of reducing cost.

Starting with the stage of initial research, we have to identify that the development of a new product, in this case, a chip, has some inherent costs which materialise from the very beginning. Such inherent costs as the cost of employment, the cost of basic infrastructure and the costs of any materials and equipment that may be needed, already present themselves.

Although it can be argued that these costs are not really significant towards the overall cost of the product and especially regarding the anticipated profit to be made from introducing the final product to the market, it is exactly at this stage that even such small costs may drive the overall project into failure or abandonment, because investments are urgently needed to support it. In the latter stages of the product's development, depending on its results, it may be easier to attract investments, while at the very beginning of the project, it is usually really hard to attract initial capital to cover the costs. However, at this very early stage, it is also easy to abandon the project without having incurred significant debt. There are different ways and means to attract capital in the semiconductor industry, but most of them focus on whether the product to be developed will really fill in some gap in the market and its production is really critical as well as on issues of timely development and the product's right placement in the market.[13]

As research proceeds and concrete designs are formed, costs related to manpower, equipment, tools, materials and infrastructure raise higher and higher, especially in a field like integrated circuit design where expert knowledge and experience are needed to produce a cutting edge technology product. However, as prototypes begin to be developed, the investors become slowly more and more committed to the project and investment should become gradually more stable. It is however estimated that the salary of an integrated circuits designer is around 45,000 euros per year[14][15] and such a project requires a significant number of different experts with long experience on designing and developing integrated circuits. Furthermore, equipment and materials needed may also cost some hundreds of thousands of euros or more,[16][17] while a full laboratory could cost millions of euros to built and fully equip.[16][17] Moreover, the cost of manpower increases as experts gain more experience over the years and their salaries increase. It also takes some years to develop a product, which means that infrastructure and constant workforce are needed for it during all this period of time.

Subsequently, the design candidates for the final product need to be adequately tested and evaluated. This obviously increases the cost of development as more people need to be hired and more equipment may be needed. However, as long as not all prototype candidates effectively fail this testing phase, the project is in a good standing and may become particularly attractive for investment, as it has now demonstrated some significant potential for its concept. Testing may not introduce outstanding costs in terms of equipment; it does, however, introduce costs in terms of time, workforce, and potentially infrastructure.

Testing may introduce a significant delay on the project, but it will also prove the quality and advantages of each prototype, while revealing any flaws in the design candidates, which could either completely invalidate them or introduce further costs in order to be fixed. Therefore, this may not only be a time-consuming or costly stage in itself, but it can also potentially wreck the project, if all design prototypes fail the testing and the product needs to be re-designed from scratch, which would practically mean doubling the cost of the whole research and development phase.

Finally, the research and development phase concludes with focusing development on the designs that have been selected as more optimal. This will lead into selecting a single design to be put in mass production after testing them again. Then, finally all development efforts will focus on this design and how to efficiently deal with any flaws this may have revealed during the previous stages of testing. This design, after being tested exhaustively, will hopefully be cleared for mass production. This stage doesn't seem to introduce any major new costs, but again requires time, workforce, infrastructure and some materials.

All in all, the overall costs for the research and development phase may rise well into the region of hundreds of millions of euros.[18] Furthermore, in the case of chips for smart cards, the plastic container of the chip, the body of the card, also has a certain amount of costs regarding related research, development and production, which may regard its durability, flexibility and endurance, as well as the way the chip will be integrated in it, or the efficient and correct placement of additional features such as antennas.[19][20] It has also been noted that the cost of research and development for semiconductor products exceeds that of most other high technology industries.[21]

After this stage, the selected prototype will enter mass production. Mass production does involve significant new costs, but it's also the step that will deliver the final product to be introduced into the market. The most significant costs associated with production are the costs of equipment, materials, and to a lesser extent infrastructure and manpower.

Especially the costs of equipment, material and infrastructure have driven a significant number of companies involved in the development of integrated circuits to not own their own semiconductor fabrication plants (commonly referred to as fab's), and to outsource the actual manufacturing of their chips. It should be stated that the cost of a fab is estimated to be at least a few billion euros.[16][17][22][23][24] This not only means that building such a plant is a huge investment that needs to somehow be evened out within a set time period, but also that just using such a facility has an inherent cost. It has, however, been noted that fab-less companies may a have high-growth potential, exactly because they are not burdened by the huge overhead costs associated with the construction and operation of fab's.[25]

Significant costs for the development of a chip are associated with the cost of materials for mass production. Semiconductor products are built in fab's on the surface of wafers, discs of pure silicon with a diameter of hundreds of millimetres. A series of identical chips are built, or developed, on the surface of a single wafer side by side, using a series of chemical, light and electrical effects. Each one of these integrated circuits developed on the surface of a wafer will be dissected from the wafer and form a single fully functional unit, called *die*.

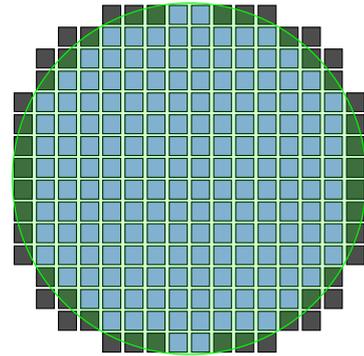

*Picture 1: Schematic representation of dice on a wafer; blue ones can be fully developed on the wafer, while green/black ones cannot and will be thus rejected, as their black area falls outside the wafer's surface.*[29]

Different sizes of wafers exist, ranging from 100, or less, to 450 millimetres. The most commonly used wafer size is 300 mm, with 200 mm wafers being phased out and 450 mm ones being slowly introduced. Bigger wafers would obviously result in the production of more chips (dice) per wafer, but the introduction of a bigger size of wafers undoubtedly would introduce also a large amount of transition costs associated with infrastructure and equipment for fab's. However, the semiconductors fabrication industry is slowly moving towards adopting an updated standard of 450mm wafers for the production of new integrated circuits.[26][27][28]

At present, most chips, including the ones used in smart cards, are developed using 300 mm wafers. Chips made to be used in smart cards are developed with a technology process of 90 nanometres, which means that each transistor contained in the chips will be approximately 90 nm wide. Each such chip usually has an area of around 10 mm² [12], which allows us to calculate the number of chips (dice) per wafer, taking into account also the circular shape of the wafer.

After the dice have been fully developed, they are extensively tested to determine whether they function properly. The proportion of the dice found to perform properly is referred to as the wafer's yield.[30]

The following formulas apply:[31][32][33]

$$\text{total cost for a chip} = \frac{\text{die cost} + \text{testing cost} + \text{packaging cost}}{\text{final test yield}}$$

$$\text{die cost} = \frac{\text{wafer cost}}{\text{dice per wafer} * \text{die yield}}$$

$$\text{dice per wafer} = \frac{\pi * \left(\frac{\text{wafer diameter}}{2}\right)^2}{\text{die area}} - \frac{\pi * \text{wafer diameter}}{\sqrt{2} * \text{die area}} - \text{dice used for testing per wafer}$$

$$\text{die yield} = \text{wafer yield} * \left(1 + \frac{(\text{defects per unit area} * \text{die area})}{a}\right)^{-a}$$

$$a = \text{manufacturing complexity (number of masking levels)}$$

It is also estimated that a 300 mm wafer costs around 2000-2500 euros.[34][35][36][37] Therefore, assuming:

- that at the stage of mass production no dice are produced only for testing any more,
- that each wafer costs around 2250 euros,
- that the area of a die is around 0.1 mm²,[12]
- that wafer yield is around 90%,
- that defects per unit area are around 10%,
- and, that we use 4 masking levels,

we get the following results:

$$\text{die yield} = 0.9 * \left(1 + \frac{(0.1 * 0.00001)}{4}\right)^{-4} \approx 0.8999991$$

$$\text{dice per wafer} = \frac{\pi * \left(\frac{0.3}{2}\right)^2}{0.00001} - \frac{\pi * 0.3}{\sqrt{2} * 0.00001} - 0 \approx 6857$$

$$\text{die cost} = \frac{2250}{6857 * 0.8999991} \approx 0.364591 \text{ euros} \approx 36.5 \text{ euro cents}$$

Also, assuming:

- a testing cost of 0.15 euro per die,[37]
- and, a packaging cost of 0.08 euro per die,[37]
- and a final test yield of 99.2%,[37]

we get the following result:

$$\text{total cost for a chip} = \frac{0.365 + 0.15 + 0.08}{0.992} \approx 0.5997984 \approx 0.6 \text{ euros} = 60 \text{ euro cents}$$

Therefore, I conclude that each chip costs around 60 euro cents to be produced in conditions of mass production, while more than 6100[1] chips are produced from each wafer.

---

1  dice per wafer * die yield * final test yield ≈ 6121

A typical fab produces 5000 to 8000 wafers per week,[38] which means that around 30,500,000 to 48,800,000 chips can be produced every week from a single fab.

It can also be concluded that chips produced for testing and/or as prototypes in the previous stages of chip development would cost quite more than chips produced in mass production. However, the little number of such testing and prototype chips can allow us to assume that such costs are fully included in the overall sum of research and development costs.

The final stage of the chip's development has to do with its introduction to the market, and can include costs related to marketing, sales and taxes. Other related costs may have to do with equipment maintenance, small improvements that may be done to the chips design and/or optimisation of the development process in general and improvements of its quality. Such costs should probably account only for a small proportion of the costs of the other stages of the chip's production, especially considering the high cost of the research and development phase. One way to calculate the final price of a chip is to classify costs in the following three main categories:[31][32]

- Component cost, having to do with the actual materials and packaging.
- Direct cost, related to labour, purchases and general recurring costs.
- Gross margin, concerning non-recurring costs, such as R&D, marketing, sales, equipment maintenance, rentals, financing and the initial investment and infrastructure.

In addition to these three categories, also an average discount category is suggested to contribute to the final list price. This category does not represent any actual costs, but makes up for the volume discounts and/or retailer mark-ups, which will reduce the list price into the actual average selling price.[31][32]

# The effects of integrating security features in chips on the previously identified agents of cost

Integrating security features in chips designed for smart cards can have immense effects on their cost. A higher initial investment will be required to cover the costs of increased manpower and infrastructure needed, while more capital will be demanded to pay for costs of additional design and testing of these features.

Even from the first phase of initial research and development of the chip, additional costs will occur, being related to (re-)designing the chip's structure and architecture, considering the new security features that have to be added. These features will not only have to be designed on their own, but also be placed in an adequate way among the other components of the chip. This will lead into more personnel and equipment being needed to provide for these additional designs, along with more materials and general infrastructure.

Moreover, such security measures implemented on chips may also require more memory and other peripheral circuits to work adequately. It has been stated that custom logic components demand not only costs for their own circuitry, but often may also introduce additional challenges regarding the overall chip, such as connection restrictions between transistors, stringent requirements on signal arrival times, or precharge signal distribution to each transistor.[39]

Furthermore, more prototypes and testing of these will be required in order to identify the best way to incorporate such security features in the rest of the integrated circuit. In addition to prototypes being required for these security modules and for the overall chips where they have been integrated, these newly designed security modules will have to be tested extensively. Freshly designed, or redesigned, security components will need to be tested on their own, while, their integration on the chip and their operation regarding the other chip's modules will also have to be tested. Finally, the whole chip's functions will have to be tested as well.

Apart from the usual testing of their functionality and general well-being, such features will also have to be tested on the security they may actually be providing. This may result into changes being needed in their designs and/or in the way they are placed on the chip. All these additional designs, prototypes and testing will obviously result into also more time being required for the best implementation to be identified and sent out for mass production. In this case, as time means more costs regarding employment, equipment, materials and general infrastructure, this results into a significant increase in cost required for this stage of development. It can be estimated that, in the best case, at least some additional months, if not much more, will be needed in comparison to the time required for the research and development of a normal chip.

As stated before, the overall costs for research and development of a normal chip are in the region of hundreds of millions of euros, while this phase may last some years. An additional amount of costs being required per year, together with an increase in the time needed for this phase, may also result in a growth of cost by a significant additional amount of money, which could be in the region of millions of euros.

In addition to this, more area will be needed for the complete chip to be produced during the mass production phase, while also issues related to the power needed by it during its operation, as well as for its production, may also arise.[40] Furthermore, such chips may have increased work loads and require more computation time than normal ones to perform the same processes.[40]

It has been calculated that each additional set of security features that may be added on a chip could, in average, add up to 10% more area to it,[41] thus requiring a larger die for its production. Depending on the initial chip size and the security features that may be added, this increase in size may differ, but it can be suggested that for such proposed systems to be feasible and, perhaps, economically viable, the increase in size should not be above 10% in general.[41][42] Moreover, the power consumption should also not increase too dramatically.[42][43] And, finally, time delays caused by the security features added in the chip should additionally be minimised as much as possible.[43]

Disregarding the costs related to the chip's performance and focusing only on a potential increase in its size, we can get the following results from the previous formulas regarding its cost, assuming a 10% increase of its size and that all previous assumptions still hold true:

$$die\ yield = 0.9 * \left(1 + \frac{(0.1 * 0.000011)}{4}\right)^{-4} \approx 0.89999901$$

$$dice\ per\ wafer = \frac{\pi * \left(\frac{0.3}{2}\right)^2}{0.000011} - \frac{\pi * 0.3}{\sqrt{2 * 0.000011}} - 0 \approx 6225$$

$$die\ cost = \frac{2250}{6225 * 0.89999901} \approx 0.401607\ euros \approx 40.2\ euro\ cents$$

$$total\ cost\ for\ a\ chip = \frac{0.402 + 0.15 + 0.08}{0.992} \approx 0.6370968 \approx 0.64\ euros = 64\ euro\ cents$$

Therefore, it can be concluded that a 10% increase in the area of the die, in order for it to incorporate additional security measures, would result in a 6.67% increase in its total production cost. This means that such a small increase in area caused 2/3 of its magnitude to appear as an increase of the production cost. In addition to this, larger chips with more functions, obviously, lead to higher testing costs after their production, as they have to be proven fully functional.

However, like before, we can also calculate that more than 5500[2] chips would be produced from each wafer. Since a typical fab produces 5000 to 8000 wafers per week,[38] around 27,500,000 to 44,000,000 secure chips could be produced every week from a single fab.

We can also assume that marketing costs may be increased or that increased taxes may apply. However, increased sales should also be expected as there is already a thriving market for secure chips, especially in the smart cards sector. We should not expect significant changes regarding costs for equipment or infrastructure, as it is an inherent requirement for such security features to not cause significant changes in the line and process of production, exactly in order to keep cost within acceptable levels.[44] As a fab costs some billions of euros to be fully productive, any changes in its equipment and/or production line can be assumed to carry significant additional costs.

It has been noted that driving the overall costs down significantly influences whether a new product can enter the market at a competitive price.[45] Additionally, it has also been suggested that even though security technology may come at a cost, network effects could also influence its initial deployment, with its benefits sometimes depending on the number of users who adopt it.[46] Costs may exceed benefits until a certain number of users adopt it, which also may be dependent on economies of scale. This, in turn, could potentially lead into everyone waiting for others to adopt first, and therefore the new technology may actually never get deployed.[46]

---

2   dice per wafer * die yield * final test yield ≈ 5557

However, it has to be noted that the secure integrated circuit industry is a thriving one and its products don't have to be cheap, as they incorporate a significant degree of uniqueness;[31] they should however demonstrate sufficient balance between providing adequate security and keeping development and production costs under control. Finally, adding security features causes such a chip to enter the market later than a normal one, which could also affect its placement and position in the market.

# The cost of performing attacks against secure chips and an estimation of potential damages and losses caused by them

As it has been mentioned before, hardware attacks can cause damages or losses amounting to several millions of euros. Especially when chips integrated in bank cards are targeted, the losses may potentially be in the region of billions of euros.[47] However, attacks, much like the actual production of such chips, come at a certain cost. Different factors depending on the chip that is being targeted can cause this cost to significantly rise.

The actual reason why an attack against a secure chip can potentially have a high cost is because it may essentially follow fully the stages of the chip's development and production process in a reverse order. Moreover, even the same or similar equipment is used in both cases, with the attacker wanting to essentially break the integrated circuit's embedded design and logic down into simple pieces, which sometimes may mean actually dissecting the chip layer by layer.

Therefore, it is easy to understand that costs related to equipment, manpower and infrastructure apply for attacks, as well. Furthermore, designing a hardware attack also requires a long period of time and a large set of different methods to be tested. This carries an inherent cost for the acquisition of materials, which in this case, are no longer raw silicon or wafers, but actual chips, which may not just be too expensive to buy in bulk, but may not even be available for purchase by the public. It is thus really important that the attacker makes the best out of the scarce materials he has possession on.

It is exactly this scarcity of materials that may make the attack worthwhile, as chips that are in wide circulation will usually not provide as much benefits when successfully compromised. Furthermore, it is this scarcity that again introduces another additional cost, which is related to expert knowledge; one or more experts have to be employed for at least some period of time if the attack is to have an adequate probability of succeeding. This means that mounting an attack against a secure chip may well come at a fraction of its research and development cost.

Moreover, an attack against a secure integrated circuit does not have a certain result, it may well keep failing for a large period of time or fail altogether. Therefore, there is always an increased risk of investment and, as the attack will usually be illegal, there are very limited possibilities for raising adequate funds to reach success. It is for this reason that usually attacks are first developed either by academic researchers or by organised criminals, competitors, or intelligence agencies. Not everyone who may develop an attack may fall into such a category, but it is those categories that can easily acquire the needed equipment, infrastructure, manpower and funds to perform a successful hardware attack. Of course there will always be people successful in developing attacks against secure chips on their own, but, in most cases, expert knowledge as well as adequate capital to cover the costs of such attacks are really required.[11]

The main characteristic which differentiates the cost of hardware attacks is their level of difficulty. This level, in turn, is based on the actual security of the chip. This security should not address just a single attack vector, but try to cover as many of them as possible. Otherwise, a chip that may be completely vulnerable to even the most expensive attacks of one kind, may be completely vulnerable to really cheap attacks of another kind, which of course would make attacking it an overall really cheap matter. In order to prevent this, a balance between security and the cost of potential attacks must be achieved, while the balance between production costs and security is also maintained.

This is really critical for secure chips and is also recognised by their classifiers, the certification and accreditation authorities, which test secure chips against a very broad spectrum of possible attacks.[48][49][50] Furthermore, it was quite early recognised that hardware can never become invulnerable to each and every kind of attack.[51] This led to the broad acceptance of the idea that security should not aim to prevent the unavoidable penetration of secure chips, but to try to postpone this and make it as difficult as possible. Therefore, (hardware) security is now defined as a state, not completely devoid of risks, but free from *unacceptable* risk.[48]

To this end, a way of lowering risk to an acceptable level is to either make attacks against a chip way too expensive in a general sense, or too expensive in comparison to their potential benefits. If an integrated circuit is vulnerable only against an attack that would cost hundreds of millions of euros to employ successfully, then, this chip can be both considered in general and classified by a certification authority as extremely secure.

Cost is therefore a critical factor for attacks targeted at secure hardware, and it is really essential to analyse the factors that produce it in respect to conducting such an attack. Thus, based on what I have already mentioned, we can again estimate that the average salary for a professional expert may be around 45,000 euros per year, depending on each person's skills and experience. However, a researcher's actual salary may significantly vary depending on his actual employer, which in this case could be an academic institution, a government, or, even, a criminal organisation.

Furthermore, different sets of tools and equipment may be needed, depending on how hard it is to penetrate the security measures of the chip and/or what kind of attack has been chosen to be tried. For different levels of actual physical penetration, different costs have been estimated, ranging from 75,000 to 370,000 euros,[11] with several different methods and equipment needed being listed in relevant publications,[16] which may range from 30,000 euros for laser cutters to between 370,000 and 730,000 thousands for different kinds of electron microscopy and spectroscopy. Thus, a full laboratory could again cost millions of euros to build and fully equip.[16][17]

Again infrastructure costs apply, but this time as this kind of activity may or may not be legal, they may have to be higher, if this is criminal activity, or lower, if it is pure academic research. The main costs however are all related to the time needed to achieve a successful attack. Furthermore, there's an inherent cost for the acquisition of chips to be attacked, which may not be available or come at a really high price per unit.

Moreover, we must distinguish between different attacks at different stages of the chip's development and production, as culprits may try to alter the chip's initial design introducing vulnerabilities or back doors into it while it is being developed or manufactured, rather than just attack it when it has already been securely produced.[49][50][52] Such mechanisms that will make the chip more vulnerable to attacks through its own design are called trojan horses, or trojans, and can be prevented by ascertaining that all stages of a chip's production are secure.[49][50][52]

However, this is not really feasible, unless all stages of development and production are being done by the same company and under heavy scrutiny and protection. In the case that the chip has to be mass produced by a third-party fab or third-party tools, equipment or design software is used, then the whole notion of secure development and production is merely based on privacy and trust. Nevertheless, the same principles of trust and privacy apply during the certification and accreditation processes of the chip, when its design, development and production details are revealed to a third-party, the certifying authority. And, of course, the same principles apply for each individual employee who may just breach his/her non-disclosure agreements and reveal critical information about the chip's security.

Therefore, even though a massively produced secure chip's design is quite static, its level of security has its own dynamics, which are far from static. To this end, the United States (U.S.) government has had a program by which a fab can be rated as a "Trusted Foundry" under stringent criteria.[53] However, as this program ran under the U.S. Department of Defence and its National Security Agency (NSA), one may not be sure if actually these fab's are really trustworthy and secure, or factories implanting trojan horses under U.S. control. As security is based heavily on trust, it is really more difficult for any product or process to be considered as truly secure after Edward Snowden's exposure of the spying activities of the U.S. NSA.

## The case of bank cards

Considering chips integrated on bank cards, in particular, we can immediately identify their abundance and relative ease of getting access into them, by simply applying for such cards. Therefore, in this case, their security cannot really be based on the unavailability of their design or the difficulty of actually getting possession of their chips. After all, most banks are only too happy to replace a "lost" bank card with a new one, which could lead to a duplication of available materials within some days.

Thus, for these and other reasons, banks have adopted other additional security measures, such as the introduction of security codes, pictures and holograms printed on the cards, magnetic strips and/or using the card holder's name and signature. A very well-known example of such a security code is the card's PIN (Personal Identification Number). Finally, in order to achieve security by limiting the unacceptable risk, most banks have introduced limits in the amount of transactions that a single card can do each day or the location at which these transactions can be done. This significantly limits risk by limiting potential losses from attacks to a specified amount per card.

However, this does not mean that the chip found on most bank cards is not really "secure". These chips do implement as many security measures as possible. Nevertheless, as the probabilities of their chips being compromised do rise to a level of risk that cannot be considered acceptable, their security cannot be based on only their secure chips. Multiple different security mechanisms ensure that their wide circulation will not affect significantly their security. Secure software and backend verification and logging also try to ensure that the whole banking system is as safe and secure as possible. In general, the main aim is to make a successful attack as expensive to perform as possible, while keeping its results as isolated as they can be. To this end, it is really important to ensure that a successful attack on a bank card cannot really compromise the whole banking system.[54]

While most attack attempts do not have immediate economic results, once one of them is successful, and especially if it becomes quickly well-known or is easy to implement, there can be an avalanche effect, which can lead into a very fast multiplication of losses and damages caused by this attack.[55] It is therefore essential to design an effective way to quarantine any compromised elements or the initial attack itself and limit its results to an acceptable level of unwanted effects. Hardware attacks may come at an initial cost of hundreds of thousands, or even several millions, of euros, but can also result in losses of millions, or even billions, of euros.[52]

Yet another really important effect that can cause real economic disasters is the breach of trust caused by a successful attack. This effect can cause severe damages to a secure chips manufacturer regarding its prestige, reputation and brand name and even cause it to eventually go bankrupt. It can also lead to litigation against such a manufacturer regarding claims for compensation or breaches of security clauses in contracts, which may also cost millions.

Finally, it must be noted that, in the worst case, it may be necessary to shut down the whole system and block all existing cards, while designing, manufacturing and issuing new cards immune to a particular attack, which in the case of a large system could take more than half a year,[56] and could, obviously, lead into losses of many millions, or even billions, of euros, considering the large number of cards involved and the considerable amount of time it would take to replace them. Furthermore, points of sales and automatic teller machines (ATMs) may have to be replaced too, which would add up a really significant cost and delay to the overall project.

In such a case, most companies would go bankrupt, as the costs of handling it are immensely huge and the period of time during which their services would have to be unavailable is quite significant. It would essentially mean that for six months or more, such a company would return to the years before smart cards were introduced. It is more than evident that this would be a huge disaster in financial terms, as well as in terms of public relations, trust, reputation, brand name and prestige.

# PART III

## The significance of security: Is it worth it?

Although it should be apparent by now that security in integrated circuits is really important for online, or offline, transactions, it is important that this notion is put into perspective. Security may be really unavoidable and actively sought-after, but it does come at a significant cost. We should therefore generally examine how much security is needed in relation to its costs, while also associating this relation with a different relation between security and risks, such as potential losses and damages. It has been suggested that this is a very difficult subject to be clearly defined in real terms, but, we could, nevertheless, attempt a brief examination of these relations.

For the particular case of smart card chips, I have proved that a potential addition of security features could result in additional costs of millions of euros for the research and development phase, while it also leads to additional costs for production, testing and marketing. Specifically, if the chip's area was to be increased by around 10%, that would mean that the chip's final price would be increased by around 6.67%. In general, we can conclude that adding security features to a chip comes at a significant cost of tens or hundreds of millions of euros, depending on each individual case.

On the other hand, I have also found out that an attack against a secure chip could cost, potentially a lot of, millions of euros to develop and test until it is successful and can be employed in a large scale. Of course, we cannot perform an immediate comparison between the costs of developing an attack and the costs of developing a secure chip, as these are highly dependent on each individual case. We can, however, compare the costs of developing a feasible attack with its financial goal of causing damages and/or losses of many more millions of euros than it costs to be developed. This holds true only for attacks developed for this aim and not for those developed particularly and exclusively for scientific research and relevant reasons. However, even attacks developed for scientific purposes could lead to huge damages and losses, but do not, or at least should not, aim for this.

We should have in mind that developing an attack could be somewhat comparable in terms of cost to actually developing such a chip, as, in a worst case scenario, such an attack would have to follow quite fully the stages of the chip's development and production process in a reverse order. Therefore, for such an attack to be economically feasible, its financial benefits should be much higher. This indirectly implies that potential losses and damages from a successful attack on a secure chip are potentially much higher than the cost of developing and producing such a chip. Even if that suggestion does not hold true in each and every case, we should also consider that not every chip incorporates security features and mechanisms.

In the case of a chip without any integral security mechanisms, its production costs may be quite lower than those of a secure chip, but if such a chip was to be used for secure applications, it could be easily attacked at a really low cost. Such a cost of developing an attack for a normal chip could come at some thousands of euros or, in the worst case, a couple of millions of euros, if the chip was really complex and with a large amount of circuits and functions. Essentially, though, the cost of attacking it could actually be lower than the cost of developing it, and, of course, the expected benefits would significantly surpass by a very large margin the cost of developing such an attack.

Furthermore, we should take in note the costs of dealing with a successful attack, which also act as incentives for increased security. Such costs include the costs of additional research, development and testing required for patching up the security hole in the chip's design, which will take up time, workforce, equipment, materials and infrastructure that could have been used for the development of other products. Furthermore, there are costs associated with the manufacturer's reputation, brand name and prestige which can potentially lead into losing current clients and failing to acquire enough future ones. Finally, there are costs related to potential compensation and restitution of affected clients, as well as general costs generated for society at large, such as law enforcement costs associated with such attacks and so on.[57]

Moreover, it has also been suggested that such measures as increasing the cost of attacks by increasing the associated penalties, strengthening national and international law enforcement and increasing the difficulty of publicising and/or promoting an attack to others will affect the market for hardware attacks directly, while also having repercussions on the market for security.[57] Most likely such measures can reduce the overall level of security-related costs, however it's not certain if they will increase the level of security as accepting a certain level of insecurity can be judged as economically rational.[57]

It is, therefore, quite clear that developing integral security features and mechanisms in integrated circuits is not only inherently unavoidable for chips that are used in secure applications, but it also is economically feasible and should be funded in a constant manner, so that it can always evolve and improve itself in the ongoing struggle of protecting an integrated circuit against both present and potential future attacks.

## Ways and means of reducing cost in the production of secure chips

Since it is important to drive the overall product cost down in order to bring a new product to the market at a competitive price,[45] it is really crucial to identify possible ways and means of reducing the development and production costs of secure chips. Such ways may include improving marketing to make use of network effects in the market sector of security technology, which will also provide for economies of scale, or modifying the development and production process for such chips.

Easy solutions as lowering the security of chips in order to lower their complexity, and therefore their costs, or providing less vulnerable integrated circuits at the same cost can be immediately rejected as they have already been proven to be neither financially feasible nor viable implementations. On the contrary, providing more incentives for security may lead into more products and cause the overall production costs to be lowered because of economies of scale in related markets. For example, since equipment may cost millions of euros to acquire, increased demand could help lower its price, and thus costs spent on that purpose.

Furthermore, it has been suggested that a common practice of vendors could be to start off with too little security and to dump any costs related to security on the end users,[59] and, then, when they have established a dominant position in the market, to add more security than is needed, but engineer it in a way that maximises customer lock-in.[60] Although this has been suggested for software security, it could hold true for hardware security as well, as a hardware security company could sell cheap low security chips at first and then take advantage of lock-in effects to provide really secure chips at high prices that the client has to buy.

If the client company does not buy those new more secure chips, then it will be left exposed to an increased risk of its systems being compromised. If the client decides it prefers to purchase more secure chips from another company, then, it may have to come against the manufacturing company of the previous chips, which knows all the chip's vulnerabilities and shortcomings and can abuse this knowledge, until such time as all the client's chips have been replaced. Furthermore, incompatibilities, discount prices and/or contract obligations, as well as a lack of feasible alternatives, may keep the client well locked-in.

Therefore, another way to achieve lower costs in the long term would be to try to decrease the customer lock-in by providing more flexible and compatible solutions in the market, which may not benefit the dominant firms and would lower the entry barriers, but would also lead into more competition, decreased costs and lower total prices.[58]

It has also been argued that as security-related costs rise, the market may reward security-related functionality that could reduce these costs.[57] Nevertheless, it is still not clear whether the security technology market can actually become flexible enough for this and adequately cope with lock-ins, or even distinguish between empty claims and security improvements that may actually achieve cost savings.[57] In general, the market for security technology, including its secure chips sector, still appears to be a lemon market, where the actual level of security of a product is hard to be defined in detail.[58]

However, there appears to be a market demand for security improvements, especially if these can reduce the total cost of ownership by reducing risk and potential damages or losses. There is also a market for vulnerabilities,[57] as these can effectively provide both vendors and clients with a leading edge in the market, which may amount to a significant advantage over competitors.

Incompatibilities make switching between vendors really costly and thus not financially viable, while also keeping development and production costs exceptionally high. However, compatibility and/or broad-spectrum standardisation may not be actively sought-after in the security sector because most of the market's value is based on privacy and trust, as well as novelty and innovation, which could be hurt by focusing on a single development, production or operation standard.

However, there are some ways to lower the cost of production and thus the overall costs of a secure chip, such as using a larger wafer of around 450 mm, instead of a 300 mm one. Nevertheless, this change in the production line would have large transitional costs, related to equipment, infrastructure and testing, which could even be as high as a billion euros, if we take into account that the cost of a new fab is at least a couple of billions. On the other hand, using a 450 mm wafer would significantly increase the production rate of chips and thus contribute to decreasing their costs through economies of scale. For this reason, this transition from 300 mm wafers to 450 mm ones has already slowly started to take place.[26][27][28]

If we assume a 450 mm wafer, and:

- that at the stage of mass production no dice are produced only for testing any more
- that each wafer costs around 3000 euros
- that the area of a die is around 0.1 mm² [12]
- that wafer yield is around 87.5%
- that defects per unit area are around 12.5%
- and, that we have 4 masking levels,

based on the previously stated formulas, we get the following results:

$$die\ yield = 0.875 * \left(1 + \frac{(0.125 * 0.00001)}{4}\right)^{-4} \approx 0.8749989$$

$$dice\ per\ wafer = \frac{\pi * \left(\frac{0.45}{2}\right)^2}{0.00001} - \frac{\pi * 0.45}{\sqrt{2} * 0.00001} - 0 \approx 15588$$

$$die\ cost = \frac{3000}{15588 * 0.8749989} \approx 0.2199497\ euros \approx 22\ euro\ cents$$

Furthermore, assuming:
- a testing cost of 0.2 euro per die[37]

- and, a packaging cost of 0.1 euro per die[37]
- and a final test yield of 98.5%,[37]

we get the following result:

$$total\ cost\ for\ a\ chip = \frac{0.22 + 0.2 + 0.1}{0.985} \approx 0.5279188 \approx 0.53\ euros = 53\ euro\ cents$$

Thus, I observe a significant cost reduction of 7 euro cents per chip, while more than 13400[3] chips can be produced from each wafer. Since a typical fab produces 5000 to 8000 wafers per week,[38] this means that around 67,000,000 to 107,200,000 chips can be produced every week from a single fab. This is more than double the weekly production rate of chips that can be developed on a 300 mm wafer in a single fab. However, as already mentioned before, the transition costs from one technology to the other are more than enormous, too.

Other means of reducing the development and production costs of chips include using thinner wafers, leaving less space between the dice, or changing the way dice are tested or put in their packages.[40][61] However, all these solutions don't seem to affect significantly the costs of development or production of secure chips.

Additionally, various optimisations have been suggested, regarding the area, performance or power needs of the chip, such as sharing components between different security sensors and features in order to amortise their integration costs.[43] Finally, completely different design approaches such as 3D chips have also been suggested, but they can even double the chip's area and lower performance (introducing significant delays), while increasing power consumption.[62] Unfortunately, it so far appears that increased costs are the price we have to pay for security.[62]

# Conclusions

It is pretty evident from what I have already mentioned that development and production of chips for smart cards come at high costs. In particular, apart from infrastructure and materials and workforce, also high tech equipment as well as expert knowledge and significant experience are required. This raises significantly the investment cost in this market area, thus creating a significant barrier not only in terms of initial entrance in it, but also regarding sustainable operation in this field. Especially if a company wants to also fully manufacture its own chips under conditions of mass production the initial investment costs rise in the area of several billions of euros.

Moreover, it is also clear that integrating security features in such chips raises production and development costs even further. Less chips can be produced in the same time and their development and production cost more. The integration of such features may cause additional costs comparable to the original costs of production and development for a normal chip.

---

3   dice per wafer * die yield * final test yield ≈ 13434

Furthermore, by examining the costs of developing an attack I found that these may run as high as the costs of developing a chip, but they can also provide significant benefits if such an attack is successful. Additionally, security is not about making an integrated circuit fully impenetrable, as this may not even be possible, but about making a potential attack unprofitable and lowering the risk of a successful attack to an acceptable level. Therefore, I concluded that security and potential damages or losses caused by an attack are strongly related.

Furthermore, we tried to identify potential ways and means of reducing the cost of adding security in chips, but we determined that most of them involve significant costs for their deployment or are not yet feasible. It is therefore significant to again note that although we do need security, at present, there is no efficient solution towards significantly reducing the additional costs that it introduces.[50] However, the constant progression of technology could slowly drive costs down, by achieving higher production rates and removing lock-ins caused by incompatibilities and abuses of the vendors' dominating position in the market.

Finally, by comparing the costs of integrating security in chips to the potential losses and damages caused by a successful attack against them, I concluded that security is highly needed and should inherently be sought-after. Additionally, I also identified that clear relations exist not only between the integration of security in chips and its cost, but also between the level of required security in an integrated circuit and the level of risk related to it. Using this relations, we can eventually determine what level of security is required to be integrated in a particular chip and if this is also financially viable.